\documentclass[a4paper,11pt]{article}
\usepackage{pos}
\usepackage{stackrel}

\title{Cascade topologies
in rare charm decays
\& implications for CP violation}

\author*[a]{Eleftheria Solomonidi}

\affiliation[a]{\it \small Departament de F\'{i}sica Te\`{o}rica, Instituto de F\'{i}sica Corpuscular,

Universitat de Val\`encia -- Consejo Superior de Investigaciones Cient\'{i}ficas,

Parc Cient\'{i}fic, Catedr\'{a}tico Jos\'{e} Beltr\'{a}n 2, E-46980 Paterna, Valencia, Spain}

\emailAdd{elefsol@ific.uv.es}

\abstract{The CP violation observed in the hadronic decays of charmed mesons remains a puzzling open question for theorists. Calculations relying on the assumption of inelastic final-state interactions occurring between the pairs of pions and kaons fall short of the experimental value. It has been pointed out that a third channel of four pions can leave imprints on the CP asymmetries of the two-body decays. At the same time, plenty of data are available for the $4\pi$ decays of charmed mesons, as well as for the rare decays $D^0\to\pi^+\pi^-\ell^+\ell^-$. With this motivation, we study the cascade topology $D^0\to a_1(1260)^+(\to \rho(770)^0\pi^+)\,\pi^-$
, which has been measured to contribute significantly to the $4\pi$ decays, and estimate its effect on the branching ratio of the rare decays. We also explore the possibility of this topology contributing to the decay amplitude of $D^0\to\pi^+\pi^-$ and by extension to the related CP asymmetry.
}

\FullConference{9th Symposium on Prospects in the Physics of Discrete Symmetries (DISCRETE2024)\\
 2–6 Dec 2024\\
Ljubljana, Slovenia\\}

\begin{document}
\maketitle

\section{Introduction}

Flavour physics plays a pivotal role in tests of the Standard Model (SM) and indirect searches for New Physics (NP). While the theoretical and experimental programme is at a remarkably advanced level for bottom-meson processes, the charm counterpart is at an earlier stage of development. However the investigation of charm-decay phenomena is crucial and unique, as charm is the only decaying up-type quark that is bound in hadrons. In particular, charm processes are excellent candidates for NP effects to be unveiled, as they enjoy a very effective Glashow-Iliopoulos-Maiani (GIM) mechanism because of the lightness of the down-type quarks that appear in the loops. 

Currently the most intriguing experimental measurement is the CP violation that has been clearly observed in the hadronic decay $D^0\to\pi^+\pi^-$ \cite{lhcbDeltaACP, lhcbKK}. The existing theoretical calculations point to a value for the direct CP asymmetry that is much smaller than the experimental one. However, the implemented approaches may require further investigation: in particular, the method of \cite{Khodjamirian, MariaLaura} is based on the framework of light-cone sum rules, for which more tests in the charm sector would be desirable. On the other hand, the data-driven approach \cite{usACP} considers hadronic final-state interactions and is based on the assumption that the pion-pair final state rescatters predominantly to a pair of kaons. This assumption can be challenged in the context of other decay environments \cite{Kubis}, where the channel of four pions appears to be mixing sizeably with the two pions at energies close to the mass of the $D$ mesons. While the incorporation of the four-pion channel into the fully data-driven approach is currently unfeasible, it is crucial to understand the dynamics of this charm-meson decay mode. A description of those weak decays accounting for non-perturbative QCD effects can be achieved by considering a number of intermediate states comprised of strongly decaying resonances. As some of the appearing resonances have been extensively studied in effective theories and models, strong phases can be incorporated through their (data-driven) line shapes. Recent amplitude analyses of the decays to four pions as well as to two pions and two kaons \cite{CLEO, BES} explore this approach and provide some enlightening results. 

The rare decays to light hadrons and two charged leptons have also received attention in the last years. On the experimental front, the recent analyses \cite{LHCbpipimumuBr, LHCbangobs, LHCbpipiee, LHCbLambdac} have provided an unprecedent volume of information. On the theory side, it is established that the semileptonic operators are very suppressed in the charm decays, namely $C_9$ is about 10 times smaller than the equivalent coefficient in bottom decays, while $C_{10}$ vanishes at order $G_F\cdot \alpha$ \cite{deBoer}. Therefore, the decay rate overwhelmingly stems from non-local insertions of four-quark operators, in association with the electromagnetic hamiltonians of quark and lepton currents. Another important consequence of the suppression of the local operators is that a number of angular observables which require a non-zero $C_{10}$ vanish in the SM. While this property is very useful in the search for NP, until a clear signal is experimentally observed the calculation of the long-distance component of those null-test observables, which boosts the effects from NP, remains indispensable for setting meaningful bounds on the NP-driven Wilson coefficients. 

Past phenomenological works \cite{Cappiello, Gudrun-original, Bharucha, usRare} mainly consider the dilepton pair to be produced from the electromagnetic decay of a vector resonance. The calculations of $D^0\to\pi^+\pi^-\ell^+\ell^-$ also model the production of the dihadron pair via the mediation of a vector ($\rho(770)^0\equiv \rho^0$ or $\omega(782)\equiv \omega$) or scalar ($\sigma=f_0(500)$) resonance which decays strongly. Specifically, they express the decay amplitudes via intermediate quasi-two-body (Q2B) topologies, whereby the charm meson decays weakly to the two intermediate-state resonances that subsequently decay to the final-state particles. An additional normalisation factor and a constant phase are assigned to each decay chain so as to encapsulate possible further QCD effects. This approach is the same as the isobar model that is implemented in the experimental amplitude analyses \cite{CLEO, BES}. The theoretical calculations within the SM so far succeed at giving an adequate qualitative description of the process when directly compared to the experimental data \cite{usRare, Gudrun-new}. However, there are some considerable tensions both in terms of the decay rate distributions as well as the CP-symmetric angular observables. While some of the tensions might be attributed to experimental shortcomings,\footnote{See relevant comments and footnote 12 in \cite{usRare}.} the most likely explanation behind systematic deviations in the distribution over the dipion mass is the presence of some theoretically unaccounted-for contributions. Additionally, the angular observables that identically vanish in the theoretical model (irrespective of any potential NP) present some non-zero values with a significance of a few standard deviations.

Since the intermediate states of the rare decays also appear as intermediate states of the $4\pi$ or $2\pi 2K$ decays (which are however populated by many additional combinations of resonances, as in this case scalar resonances can also produce the second hadron pair in lieu of the dilepton), it is instructive to note which decay chains dominate the hadronic decay rates. In both amplitude analyses \cite{CLEO, BES}, the decay chain $D^0\to \pi^- a_1(1260)^+(\to\pi^+\rho^0(\to\pi^+\pi^-))$, which is of the cascade-type topology, where two of the pions are produced consecutively and not at the same vertex, comes out as the largest contributor to the decay rate, with a branching fraction much larger than the chain $D^0\to\rho^0(\to\pi^+\pi^-)\rho^0(\to\pi^+\pi^-)$. This result is qualitatively expectable, as in naive factorisation the amplitude for the cascade decay is created from an insertion of the Fermi operator $Q_1$, which has a Wilson coefficient about three times larger than the coefficient of $Q_2$, which appears in all the Q2B topologies. 

Given this fact and since our SM calculation for the rare decays appears still incomplete, we are motivated to re-evaluate our approach by introducing, in addition to the existing components, the cascade topology of the $a_1(1260)^+$. As per our previous work, we appropriately  assign free normalisation factors and constant phases to be fitted to the experimental mass distributions. Interference effects are also taken into account, which stem from the presence of the cascade topology in both the $S$- and $P$- waves of the pion pair. Finally, we draw a first comparison between our results and the amplitude analysis of the hadronic decays, in an attempt to evaluate the validity of the resonance-mediated model and the universality of hadronic final-state effects in charm decays. 

\section{Framework}

For the purposes of the needed precision, due to the features of the short-distance dynamics described in the previous section, the effective Hamiltonian for the decays $c\to u \mu^+\mu^-$ can be reduced to 
\begin{equation} \label{eq:H_eff}
	\mathcal{H}_{\rm eff} = \frac{G_F}{\sqrt{2}} \left[ \, \sum^2_{i = 1} C_i (\mu) \left( \lambda_d Q_i^d + \lambda_s Q_i^s \right)  \right] + \mathrm{h.c.}
\end{equation}

\noindent where $
	\lambda_q = V^\ast_{c q} V_{u q}$, $q = d, s$, and  the operators appearing are the following:

\begin{eqnarray}\label{eq:operator_list}
	 Q_1^q = ( \overline{q} c )_{V-A} ( \overline{u} q )_{V-A} \,, \quad
	&& Q_2^q = ( \overline{q}_j c_i )_{V-A} ( \overline{u}_i q_j )_{V-A} \stackrel[]{Fierz}{=} ( \overline{u} c )_{V-A} ( \overline{q} q )_{V-A} \,, \nonumber \quad q = d, s \,,
\end{eqnarray}
where $ (V - A)_\mu = \gamma_\mu (\mathbf{1} - \gamma_5) $, $i, j$ are colour indices, and $\mu \sim \overline{m}_c (\overline{m}_c)$ is the renormalisation scale.

The $S$-matrix elements can be schematically written as follows: 

\begin{eqnarray}
     \langle \pi^+ \pi^- \ell^+ \ell^- | S |D^0 \rangle = && \langle \pi^+ \pi^- \ell^+ \ell^- |  \int d^4 x \, d^4 w \, d^4 y \, d^4 z \\
    && T \{ \mathcal{H}_{em}^{\rm lept}(z) \, \mathcal{H}_{\mathcal{V}\gamma}(y) \, \mathcal{H}_{\mathcal{R} \pi\pi}(w) \, \mathcal{H}_{D \mathcal{R} \mathcal{V}}(x) \} |D^0 \rangle \,, \nonumber
\end{eqnarray}%
for the Q2B topologies, where $\mathcal{R}=\sigma, \rho^0$ or a small isospin-violating $\omega$  component, while $\mathcal{V}=\rho^0,\omega,\phi(1020)\equiv \phi$ are the vector resonances which couple to a single photon via the electromagnetic hamiltonian $\mathcal{H}_{\mathcal{V}\gamma}$; and
\begin{eqnarray}
    \langle \pi^+ \pi^- \ell^+ \ell^- | S |D^0 \rangle = && \langle \pi^+ \pi^- \ell^+ \ell^- |  \int d^4 x \, d^4 w \, d^4 y \, d^4 z \\
    && T \{ \mathcal{H}_{em}^{\rm lept}(z) \, \mathcal{H}_{\mathcal{V}\gamma}(y) \, \mathcal{H}_{\mathcal{A} \mathcal{V}\pi}(w) \, \mathcal{H}_{D \mathcal{A} \pi}(x) \} |D^0 \rangle \,, \nonumber
\end{eqnarray}
for the cascade topologies. We only consider $\mathcal{A}=a_1(1260)^+\equiv a_1$, coupling to $\mathcal{V}\pi^+$ with $\mathcal{V}=\rho^0$. We omit the CP-conjugate decay $D^0\to a_1(1260)^-\pi^+$, which has different weak dynamics, as a naive estimate indicates a much smaller contribution to the decay rate than the decay under study. This result is further supported by the quoted values for the decay width fractions in the amplitude analyses of $D^0\to4\pi$, which is about ten times smaller for $D^0\to a_1(1260)^-\pi^+$ than for $D^0\to a_1(1260)^+\pi^-$.  Further cascade-type decays are neglected given their reported suppression in the $D\to 4\pi$ decays and the lesser known nature of the heavier axial vector resonances. 

Therefore, the only additional $S$-matrix element, with respect to previous works, contributing to the decay is 
\begin{eqnarray}
\label{eq:Smatrix-cascade}
    \langle\pi^+ \pi^- \ell^+ \ell^-| S |D^0\rangle^{(\mathrm{casc})}= 
    (2 \pi)^4 \, \delta^{(4)} (p + q - p_D) (\overline{u}_\ell\gamma_\mu v_\ell)\, \left( \lambda_d \frac{G_F}{\sqrt{2}} \,e^2  C_1(\mu)\right) \\
    \left( f_+(k^2)(p_1+2 p_2)^\mu +f_-(k^2)p_1^\mu\right) \, \frac{m_{a_1}f_{a_1} \, g_{a_1\rho\pi} }{P_{a_1}(k^2)}  
    \,
    \frac{1}{P_\rho(q^2)} 
    \,   \frac{f_\rho}{\sqrt{2}m_\rho}  
    \, B_{\mathrm{casc}}e^{i \delta_{\mathrm{casc}}}\,,  \nonumber
\end{eqnarray}
where $k^2$ is the squared momentum of the $a_1$ meson, $B_{\mathrm{casc}}$ and $\delta_{\mathrm{casc}}$ are the free normalisation factor and phase respectively, and $f_+(k^2), f_-(k^2)$ are the two $D\to\pi$ form factors. 

As the cascade topology is kinematically different from the Q2B one, it manifests distinctly in the observables commonly presented in the experimental analyses, which are defined based on the kinematical variables $p^2, q^2, \theta_h, \theta_\ell, \phi$.\footnote{For a definition of the kinematical variables see \cite{usRare}.} Namely, since the axial resonance carries the momentum of the lepton pair along with the momentum of one pion, it depends on the three variables $p^2, q^2$ and $\theta_h$. While in the $q^2$ distribution a peak is still expected around the mass of the $\rho^0$ meson, from which the lepton pair is created, there is no sharp resonant peak associated with this topology in the $p^2$ distribution. On the other hand, in the distribution over $\cos\theta_h$ a smooth peak is expected around a value that can be calculated from the kinematics, a feature which is absent in the purely Q2B-driven decays, wherein the shape of the distribution is parabolic. Depending on the size of the cascade decay amplitude, as well as of the interferences with the Q2B topologies, with respect to the total decay, the distribution will look differently. 

In this preliminary report we focus on the effect of the cascade decay on the distribution over the invariant mass of the pion pair and compare directly to the available experimental distribution. We also note that the presence of the cascade topology renders possible the non vanishing, in the SM, of the angular observables $\langle S_8\rangle$ and $\langle S_9\rangle$, which have been measured away from zero with a few-$\sigma$ significance for some $q^2$-bins.

We calculate the amplitudes from the Q2B topologies as per our previous work \cite{usRare}. For the new cascade contribution, we implement the following: we use the value $C_1(m_c)=1.22$ at next-to-leading order (NLO) in the naive dimensional regularisation (NDR) scheme, and the $D\to\pi$ form factors  from the lattice \cite{fnal-DtoPi} fitted to a nearest-pole-approximation formula. The rest of the constants appearing in the matrix element of Eq.~\eqref{eq:Smatrix-cascade} can be absorbed in the normalisation factor that is fitted to the data.

\section{Results}

The fit to the differential distribution over the dipion mass is sensitive to the following free parameters: 6 normalisation factors, namely the one for the cascade decay, those for the $\rho^0\rho^0$, $\rho^0\phi$, $\sigma\rho^0$ and $\sigma\phi$ amplitudes (the decay $D\to\rho^0\omega(\to\ell^+\ell^-)$ is suppressed \cite{usRare}, and the $p^2$ distribution is not sensitive to the size of the small $\sigma\omega$ contribution), the value of the $\omega\to\pi\pi$ admixture, $a_\omega$; and 3 phases: the phase of the admixture $\phi_\omega$ and the 2 relative phases between the cascade amplitude and the $S$- and $P$-wave Q2B amplitudes. 

\begin{figure}
    \centering
        \includegraphics[width=0.45\linewidth]{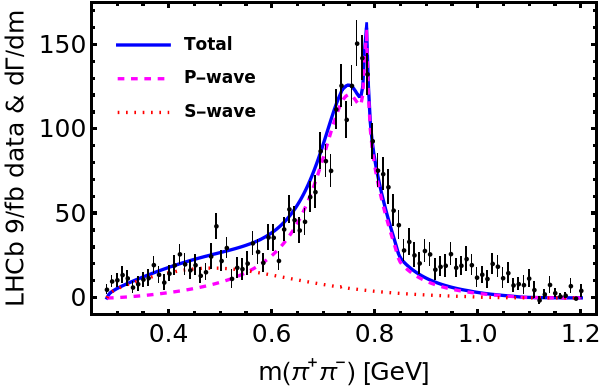}
        \includegraphics[width=0.45\linewidth]{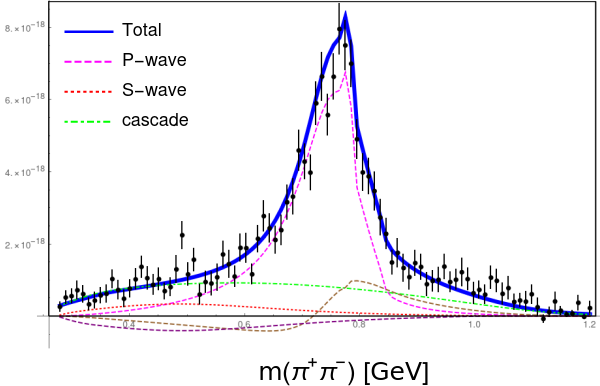}
    \caption{The prediction for the differential decay rate
$d\Gamma/dm$ and LHCb data over the di-hadron invariant
mass $m(\pi^+\pi^-) \equiv \sqrt{p^2}$: (left) as calculated in \cite{usRare} without the inclusion of the cascade topology, shown in the scale of LHCb events and with arbitrary normalisation for the theoretical prediction; (right) preliminary prediction with the inclusion of the cascade topology, shown in dotted-dashed green, as well as interference effects with the Q2B topologies, shown in dashed brown ($S$-wave) and dashed purple ($P$-wave), in the real scale of the $d\Gamma/dm$ (in GeV) both as predicted by the theory and as measured by LHCb in \cite{LHCbpipimumuBr}.}
    \label{fig:dBrdp}
\end{figure}

The results of the preliminary fit can be seen on the right panel of fig.~\ref{fig:dBrdp}. The fit describes overall well the experimental data for most of the dipion mass range, with the exceptions of the regions around the mass of $K_S^0$ and above 1 GeV, for reasons explained in \cite{usRare}. A value of $\chi^2_\mathrm{min}/\mathrm{dof}\approx 1$ is obtained. With respect to our previous result, plotted on the left panel of the same figure, a significant improvement is clearly visible, most notably above the $\rho/\omega$ peak. We also notice that the distribution from the interference between cascade and $\rho\mathcal{V}$ or $\sigma\mathcal{V}$ takes negative values for a big region of the dipion invariant mass. In the presence of the cascade component, the fit prefers a smaller S-wave Q2B contribution, which is expectable given that the shape of the purely cascade distribution resembles the one of $\sigma\mathcal{V}$  for small dipion masses. However, the inclusion of the $\sigma\mathcal{V}$ amplitude is still important, both for its interference effects with the cascade needed in order to fit the $m(\pi^+\pi^-)$ distribution, as well as for describing the rest of the observables which are not discussed in this proceeding. 

Finally, we remark that the fitted value for the normalisation parameter of the cascade topology is in excellent agreement with the value extracted from the amplitude analysis of $D\to4\pi$, an encouraging result towards a consistent description of long-distance effects across fully hadronic and rare charm decays. Further comparisons are to appear in the completed manuscript.

\section{Conclusions}

Despite the common idea that the long-distance-dominated observables are not useful in the search for NP, a good control of the hadronic effects is still very valuable especially in charm physics for constraining and potentially measuring NP which can clearly manifest in the so-called SM-null-test observables. In this work we have improved significantly the SM description of the rare $D^0\to\pi^+\pi^-\mu^+\mu^-$ decays with the inclusion of a single cascade-type decay proceeding through the axial vector resonance $a_1(1260)^+$. Furthermore, our work corroborates the finding of recent amplitude analyses that this decay chain is dominant in the weak decays of the $D^0$ meson to four pions. With our view to the CP asymmetries of fully hadronic decays, where large tensions between experiment and theory are currently present, an improvement of the theoretical prediction that is based on the inclusion of the four-pion channel interacting with the two-body final states \cite{usFuture} would then need to take into account the $a_1^+\pi^-$-mediated decay. 

\subsection*{Acknowledgements}
The author thanks Luiz Vale Silva for the collaboration throughout the project and for reading the proceeding, as well as Svjetlana Fajfer for the hospitality and the scientific insight during the months when this project took form. She also thanks Gudrun Hiller, Luka Leskovec, Dominik Mitzel, Sasa Prelovsek Komelj, Pablo Roig and Dominik Suelmann for useful discussions.  

This work has been supported by MCIN/AEI/10.13039/501100011033, grant PID2020-\\114473GB-I00, and by Generalitat Valenciana, grants PROMETEO/2021/071 and the GenT program (CIDEGENT/2021/037).

\end{document}